\documentclass[prl,superscriptaddress,twocolumn,preprintnumbers,a4,showpacs,floatfix]{revtex4}  
\usepackage{amsmath, amssymb, psfrag, tabls, dcolumn, bm, color}
\usepackage[sort&compress]{natbib}
\usepackage[pdftex]{graphicx}
\usepackage{graphicx}

\newcommand{\scite}[1]{\cite{#1}}

\begin{document}

\title{Self-passivating edge reconstructions of graphene}

\author{Pekka Koskinen\footnote{Author to whom correspondence should be addressed.}}
\email{pekka.koskinen@phys.jyu.fi}
\address{Department of Physics, NanoScience Center, 40014 University of Jyv\"askyl\"a, Finland}

\author{Sami Malola}
\address{Department of Physics, NanoScience Center, 40014 University of Jyv\"askyl\"a, Finland}

\author{Hannu H\"akkinen}
\address{Department of Physics, NanoScience Center, 40014 University of Jyv\"askyl\"a, Finland}
\address{Department of Chemistry, NanoScience Center, 40014 University of Jyv\"askyl\"a, Finland}


\begin{abstract}
Planar reconstruction patterns at the zigzag and armchair edges of graphene were investigated with density functional theory. It was unexpectedly found that the zigzag edge is metastable and a planar reconstruction spontaneously takes place at room temperature. The reconstruction changes electronic structure and self-passivates the edge with respect to adsorption of atomic hydrogen from molecular atmosphere. 
\end{abstract}

\maketitle
\hfill

Carbon is one of the most prominent elements in nature, vital for biology and life. Although macroscopic carbon has been important since ancient times\scite{katsnelson_MT_07}, only modern materials design, utilizing nanotubes\scite{ijima_nature_91, baughman_science_02}, fullerenes\scite{kroto_nature_85} and single graphene sheets\scite{meyer_nature_07}, fully attempts to use its flexible chemistry. In  applications for nanoscale  materials and devices, it is often the atomic and electronic structure of boundaries and surfaces that is responsible for mechanical, electronic and chemical properties.

Since the properties of nanomaterial depend on the precise atomic geometry, its knowledge is crucial for focused preparation of experiments and for worthy theoretical modeling. Only this enables the further development of nanoelectronic components, nanoelectromechanical devices 
and hydrogen storage materials\scite{kinaret_APL_03,baughman_science_02}, or the usage of carbon in compound designs\scite{stankovich_nature_06}. 

The importance of precise geometry is emphasized in low-dimensional systems. The strong 
correlations are known to bring up novel phenomena\scite{koskinen_PRL_07}, and such should be 
expected also for the quasi-one-dimensional edges of graphene. The edge chemistry plays even 
crucial role in the catalyzed growth of carbon 
nanotubes\scite{charlier_science_97,lee_PRL_97}. Specifically, as two-dimensional carbon has 
the honeycomb lattice, edge behaviour ultimately boils down to the properties of graphene 
edges. Hence it is relevant to explore different graphene edge geometries and their chemical 
properties beyond the standard zigzag and armchair ones.

This relevance is evident from the abundant literature. The electronic properties of graphene 
as well as carbon nanotube armchair and zigzag edges have been studied 
extensively\scite{nakada_PRB_96,kawai_PRB_00}, often in connection to nanotube 
growth\scite{charlier_science_97,hernandez_JCP_00} or the so-called electronic 
``edge states''\scite{jiang_JCP_07,gunlycke_APL_07,kobayashi_PRB_06}. There has 
been experimental and theoretical work done even on the reconstruction of graphene edges, but 
they have differed from the basic reconstruction patterns studied in this work. They have 
involved either edge roughness\scite{gunlycke_APL_07} or more dramatic folding of the edge 
into a loop\scite{jian_carbon_06}.

\begin{figure}[tb]
\begin{center}
\includegraphics[angle=0,width=8.5cm]{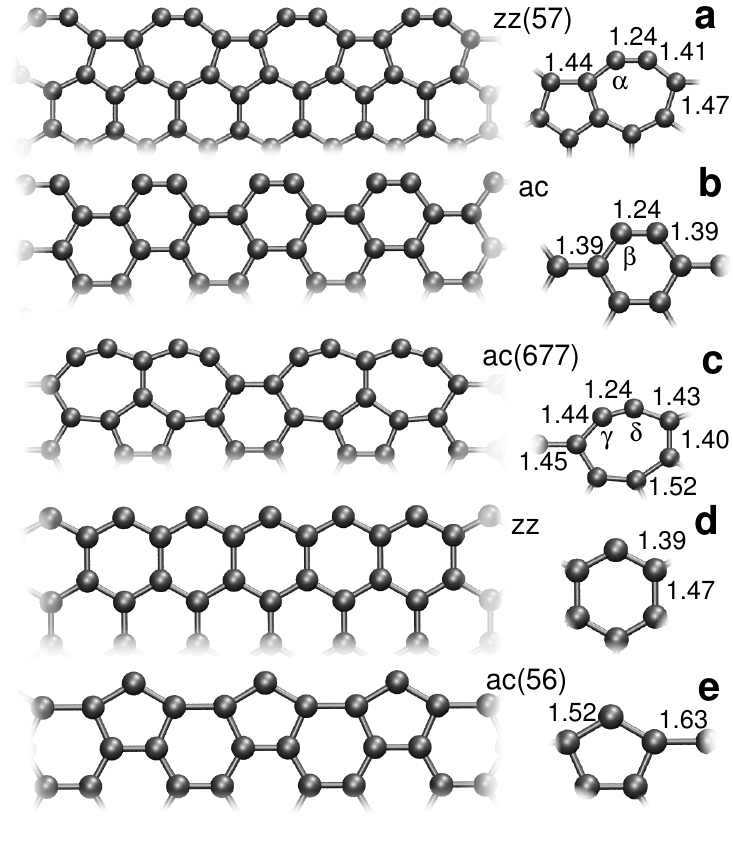}
\caption{The geometries of graphene edges. (a) reconstructed zigzag (zz(57)), (b) armchair (ac), (c) reconstructed armchair (ac(677)), (d) zigzag (zz), and (e) pentagonal armchair (ac(56)) edge. The numbers in parentheses denote the number of vertices 
in edge polygons. Some  bond lengths (in \AA) and bond angles are shown on the right: the 
bond angles are $\alpha=143^\circ$, $\beta=126^\circ$, $\gamma=148^\circ$ and 
$\delta=147^\circ$. All geometries are strictly in plane.}
\label{fig:edges}
\end{center}
\end{figure}

%
%
The edges discussed in this work are shown in figure~\ref{fig:edges}, and were investigated by modeling an infinitely long carbon nanoribbon of a given width (see Methods). Tight-binding method\scite{porezag_PRB_95} was used to explore number of other edge candidates, but density-functional analysis only for the relevant ones is reported here. The most important edge is zz(57), a reconstruction of zigzag edge where two hexagons transform into a pentagon and a heptagon, like an edge cut through a Haeckelite structure or a line of Stone-Wales defects\scite{stone_CPL_86}. The edge ac(677) is a reconstruction of the armchair edge where two separate ``armrest'' hexagons merge into adjacent heptagons by Stone-Wales mechanism. The pentagonal reconstruction ac(56) of armchair has a slightly different nature, since it requires the diffusion of carbon atoms from distant ``armrests'' to ``seat'' positions. 

\begin{figure}[tb]
\begin{center}
\includegraphics[width=8cm,angle=0]{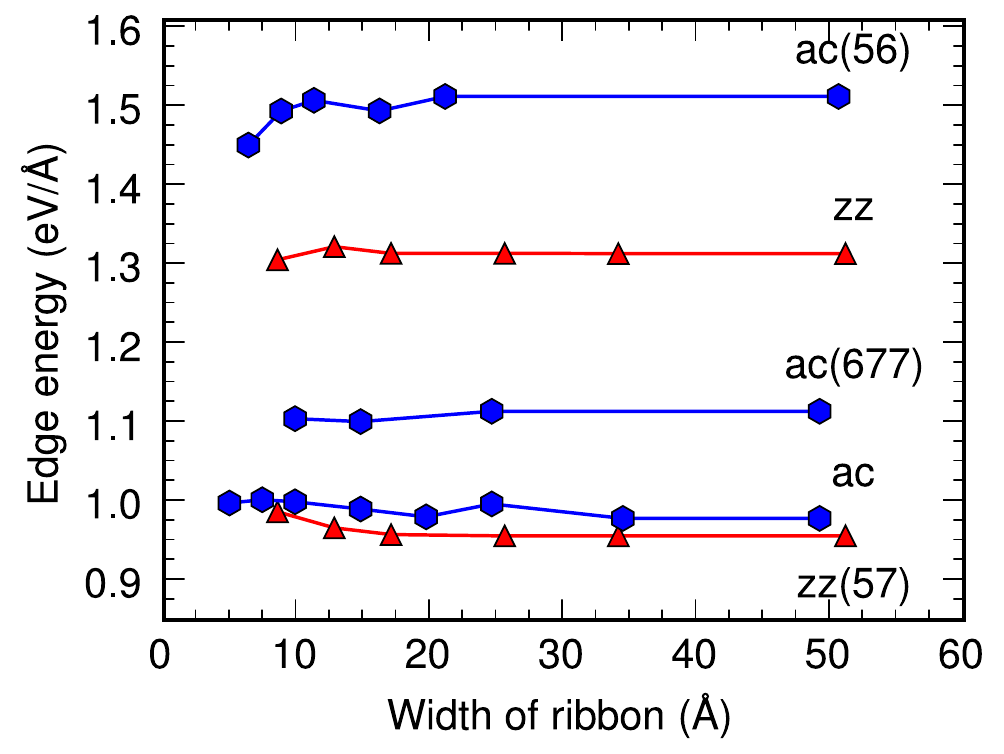}
\caption{(color online) The edge energies of carbon nanoribbons. Energies are plotted as a function of 
the ribbon width for the edges in figure~\ref{fig:edges}.}
\label{fig:width}
\end{center}
\end{figure}

Let us start analysis by looking at edge energy $\varepsilon_{edge}$, which is calculated 
from the total energy of the graphene ribbon
\[
 E=-N\cdot \varepsilon_{gr}+L \cdot\varepsilon_{edge},
\]
where $N$ is the number of carbon atoms, $L$ the total length of edges (twice the length of 
simulation cell), and $\varepsilon_{gr}=7.9$~eV is the cohesion energy of graphene. The edge 
energies converge rapidly as shown in figure~\ref{fig:width} and justify the reference to 
(semi-infinite) graphene. The energy of armchair edge is $0.33$~eV/\AA~lower than for zigzag 
edge, in accord with a similar value for nanotubes\scite{lee_PRL_97}. However, the principal 
result is that by edge reconstruction zigzag may lower its energy by $0.35$~eV/\AA. This implies that the 
\emph{reconstructed zigzag is the best edge for graphene}. To the best of our knowledge, this 
is the first report on the metastability of the zigzag edge, which is surprising in view of 
the abundant literature. Thermal stability of this novel reconstruction was also confirmed with tight-binding simulations\scite{porezag_PRB_95}. The ac(677) edge has only slightly higher energy than the armchair 
edge, whereas ac(56) reconstruction has the highest edge energy. We remark that the 
reconstructions appear to be stable with respect to out-of-plane motion, a situation somewhat 
different from small-diameter nanotubes\scite{hernandez_JCP_00,charlier_science_97}.

These energetics can be understood by looking at the geometries of figure~\ref{fig:edges}. 
Previous studies have shown that armchair has low energy due to triple bonds in the 
``armrests''\scite{kawai_PRB_00}, as realized by comparing their short bonds ($1.24$~\AA) to 
the bond in acetylene ($1.20$~\AA). Zigzag does not have such triple bonds and ends up with 
strong and expensive dangling bonds. The reconstructed zz(57) has triple bonds ($1.24$~\AA) 
but also wider bond angles ($143^\circ$) which reduces the hybridization energy cost. Triple bonds 
with wide angles can be observed also in ac(677) edge, but strain in other parts makes 
the reconstruction unfavoured. The ac(56) edge suffers from dangling bonds like zigzag, and 
additional high strain energy makes this reconstruction the most expensive one. Regardless, 
ac(56) edge has relevance during the growth of armchair edges\scite{lee_PRL_97}. 
%
%
\begin{figure}[tb]
\begin{center}
\includegraphics[width=8cm,angle=0]{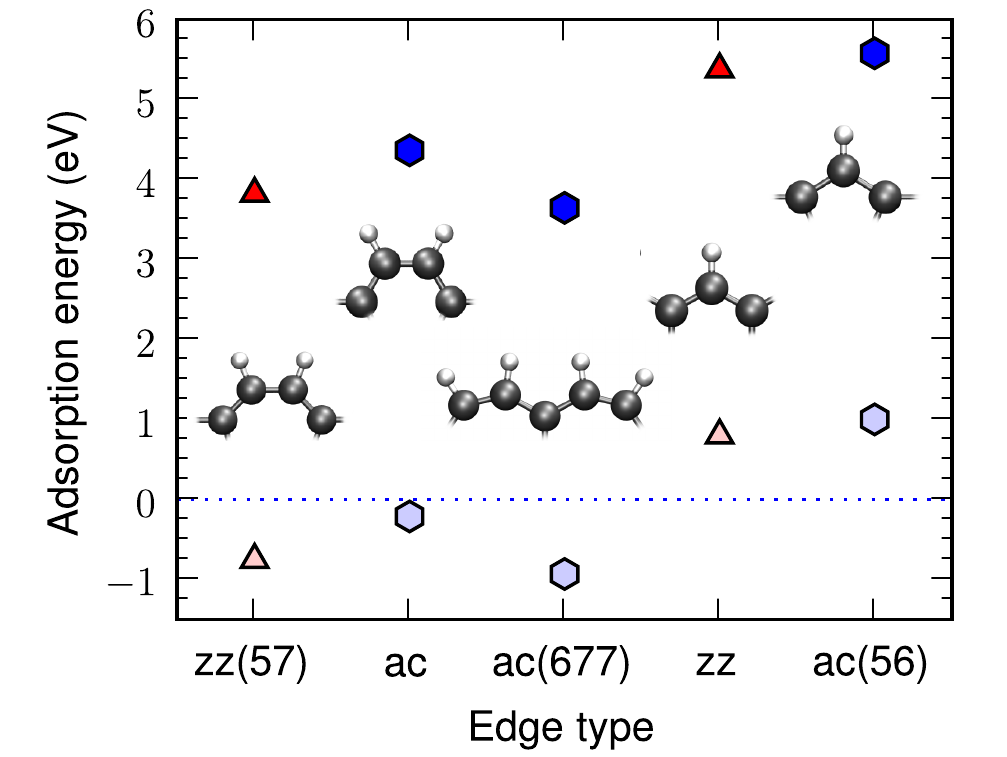}
\caption{(color online) Hydrogen adsorption energies. The upper symbols correspond to 
$\varepsilon_{ads}$ with coverage of one hydrogen per edge atom and lower faint symbols 
are shifted by subtracting the H$_2$ binding energy; positive $\varepsilon_{ads}-E_{H_2}$ 
means the hydrogen atom is more strongly bound to the edge than to H$_2$ molecule.}
\label{fig:ads}
\end{center}
\end{figure}

These observations are supported by hydrogen atom adsorption energies, shown in 
figure~\ref{fig:ads}. The weak adsorption for armchair ($4.36$~eV) compared to zigzag 
($5.36$~eV) stems from the triple bonds in the armchair edge. Similarly the weak adsorption  
for zz(57)  ($3.82$~eV)  and ac(677) ($3.64$~eV) witnesses the weakening of dangling bonds. 
The adsorption for ac(56) is large because of the strongest dangling bonds. For insight, the 
adsorption energies in figure~\ref{fig:ads} are replotted by subtracting hydrogen molecule 
binding energy $E_{H_2}=4.58$~eV. The resulting negative number 
for zz(57) means that the adsorption of hydrogen atom from H$_2$ molecule is not favored due 
to cost of H$_2$ dissociation energy, unless the adsorption process should be complicated\scite{sha_JACS_04}. This amounts to the conclusion 
that  \emph{the edge reconstruction chemically passivates zigzag edge}. However, hydrogen 
adsorption for zigzag$+$hydrogen edge has yet smaller adsorption energy of $2.14$~eV.

The edge energetics are summarized in table~\ref{tab:ads}. Note that the ordering of edges 
changes when expressed as energy per edge atom 
($\varepsilon_{edge}^*=\varepsilon_{edge}\lambda^{-1}$) due to different edge atom densities 
$\lambda_{ac}=(2.13~\textrm{\AA})^{-1}$ and $\lambda_{zz}=(2.46~\textrm{\AA})^{-1}$. More 
interesting is to look at edges with hydrogen termination (Klein 
edge)\scite{kobayashi_PRB_06}. For this case the edge energy is 
$\varepsilon_{edge+ads}=\varepsilon_{edge}-\lambda \cdot (\varepsilon_{ads}-E_{H_2}/2)$, 
where the reference is to bulk graphene and H$_2$ molecules. The best edges in this 
case are normal armchair and zigzag edges because of dangling bonds. On the contrary, the 
weak dangling bonds and small adsorption energy causes high energies for zz(57) and ac(677) 
Klein edges.

\begin{table}[tb]
\caption{Summary of the edge and hydrogen adsorption energies. $\varepsilon_{edge}^*$ is the energy per edge atom, $\varepsilon_{ads}$ the hydrogen adsorption energy with full edge coverage, and $\varepsilon_{edge+ads}$ is the edge energy with hydrogen termination. Note that for ac(56) the edge atom density is $\lambda_{ac(56)}=\lambda_{ac}/2$.}

\begin{tabular}{lccccc}
\hline
edge &zz(57)&ac&ac(677)&zz&ac(56)\\ 
\hline
$\varepsilon_{edge}$ (eV/\AA)		&0.96&0.98&1.11&1.31&1.51\\
$\varepsilon_{edge}^*$ (eV/atom)  	&2.36&2.09&2.30&3.22&6.43\\
$\varepsilon_{ads}$ (eV)			&3.82&4.36&3.64&5.36&5.58\\
$\varepsilon_{edge+ads}$ (eV/\AA)	&0.34&0.01&0.45&0.06&0.74\\
\hline
\label{tab:ads}
\end{tabular} 
\end{table}

%
%
\begin{figure}[b!]
\begin{center}
\includegraphics[width=8.5cm,angle=0]{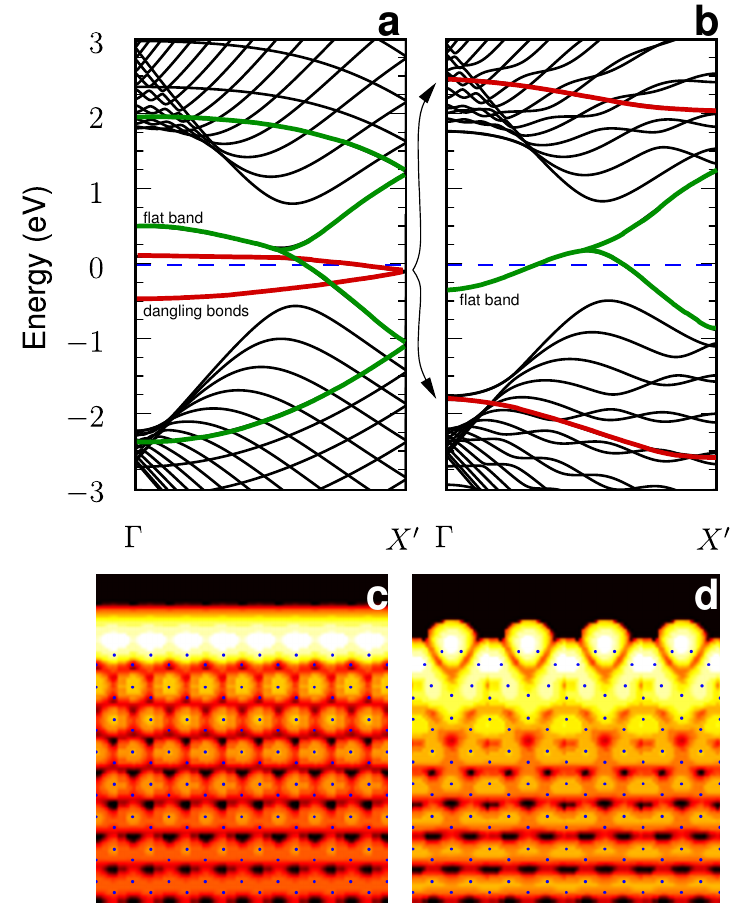}
\caption{(color) The electronic structure of zigzag and zz(57) edges. (a) and (b) shows 
the band structure for $34$~\AA~wide zigzag and zz(57) nanoribbons, respectively, with unit 
cell width of $4.9$~\AA. Note that for zigzag this is twice the minimum unit cell and the 
reciprocal space is thus only half of the normal representation. The dashed line is the 
Fermi-level. The bands coloured were identified directly by visual inspection of the wave functions. (c) and (d) show the height profiles of simulated scanning tunneling microscope images in constant current mode of 
the respective edges (height variations $>2$~\AA), formed by integrating the electron density from occupied bands within 
0.1 eV of the Fermi energy. The degeneracies at the gamma-point are 2 and 4 for the dangling 
bonds and the flat band, respectively.}
\label{fig:electr}
\end{center}
\end{figure}

%
%
Let us now concentrate on the thermodynamic and electronic properties of zigzag edges. We 
would like to clarify an aspect which is ambiguous in the literature: there are two types of 
electronic zigzag edge states. The so-called ``flat band'' in figure~\ref{fig:electr}a comes 
from the bulk $\pi$-electrons and is indeed localized at the edge. But the band due to 
dangling bonds is the one seen in the scanning tunneling microscope (STM) image and is 
located spatially even beyond the edge (figure~\ref{fig:electr}c). In zz(57) formation of 
triple bonds is evidenced by the nearly isolated dimers in figure~\ref{fig:electr}d, and the 
reconstruction removes the dangling bond bands away from the Fermi-level by lifting the 
degeneracy almost by $5$~eV. Hence for zz(57) the STM image shows only the ``flat band'' 
states. Because the dangling bond bands shift to elusive energies, also chemical reactivity 
reduces.

In thermodynamic sense the spontaneous reconstruction of zigzag into zz(57) should be 
possible, since the activation barrier from the zigzag side is only $0.6$~eV, from the 
reconstructed side $2.4$~eV. The G-mode vibration of graphene at $1580$~cm$^{-1}$ gives an 
attempt frequency of $\nu_G \sim 5\cdot 10^{12}$~s$^{-1}$, and elementary approach yields the 
rapid rate $\nu_G \cdot \exp{(-E_B/k_BT)}\approx 4\cdot 10^2$~s$^{-1}$ at room temperature. 

The reconstructions predicted in this work are expected to survive on graphite terraces due 
to the weak interaction ($5.6$~meV/atom) between the basal planes\scite{hua_nanotech_00}. 
Using appropriate sample preparation it should be thus possible to observe the 
reconstruction. STM images often show irregular and blurred edges, yielding no atomic 
resolution, but at least for passivated edges armchair predominance is claimed, in agreement 
with table~\ref{tab:ads}\scite{kobayashi_PRB_05,kobayashi_PRB_06}. So far samples have been 
prepared intentionally with hydrogen passivation during heat 
treatment\scite{fukunaga_JAC_01,kobayashi_PRB_05}, a situation where reconstruction would not be favored. 
Alternate routes for observing the reconstruction would be the radial distribution function 
from neutron diffraction experiments without deuterium atmosphere\scite{fukunaga_JAC_01}, or 
detection of triple bond -spawned high-energy modes around $2000$~cm$^{-1}$ with Raman 
spectroscopy.

A further topic is the study of the conductance of various graphene edges, particularly the 
zz(57) edge found here. The presence of the edge state around the Fermi level makes zz(57) ideal for conductance measurements, in contrast to armchair ribbons where the edge state is absent. This may render zz(57) as an interesting stable model for quasi-one-dimensional carbyne with alternating single and triple bonds. Furthermore, the novel thermodynamically and chemically stable reconstruction could play a role in formation of angular joints in nanoribbons\scite{li_science_08}, closure of the ends of nanotubes after cutting\scite{gan_NJP_08} and any other system where graphene sheets are joined to produce systems with nano-scale morphology.

\section*{Methods}
%
%
We used density-functional theory in conjunction with generalized gradient approximation for 
the exchange correlation functional\scite{perdew_PRL_96} and projector augmented 
waves\scite{blochl_PRB_94} for the C($2s2p$) electrons, as implemented in GPAW code 
using real-space grids\scite{gpaw_wiki}. Converged energies were obtained 
with grid spacing of $0.2$~\AA~ and $10$ ${\bf k}$-points in the periodic direction. In the 
perpendicular directions the system is not periodic and the space between the atoms and the 
wall of the simulation cell was $\geq 5.0$~\AA. The energies were converged to $\sim 
10^{-5}$~eV/atom and structures were optimized until forces were less than $0.05$~eV/\AA. Our 
calculations agree well with previous relevant experimental as well as theoretical energetic 
and geometric properties\scite{hua_nanotech_00,kawai_PRB_00}. The activation barrier was 
calculated with nudged elastic band method\scite{henkelman_JCP_00,bitzek_PRL_06} by fixing 
the atoms beyond the first two zigzag-rows with unit cell of length $4.9$~\AA. The constant current STM of Fig.~\ref{fig:electr} shows the height profile of electron density isosurface of the occupied states within $\sim 0.1$~eV energy window below the Fermi-level. The isosurface value corresponds to average density 2~\AA~above the graphene plane ($3\cdot 10^{-5}$~electrons/\AA$^{3}$).

\section*{Acknowledgements}
We acknowledge support from the Academy of Finland (projects 121701 and 117997) and from the 
Finnish Cultural Foundation. We thank Matti Manninen for discussions, Michael Walter 
for technical assistance with GPAW calculations and Karoliina Honkala for reading of 
the manuscript. The computational resources were provided by the Finnish IT Center for 
Science (CSC) in Espoo.

\bibliographystyle{unsrt}

%

\begin{thebibliography}{30}
\expandafter\ifx\csname natexlab\endcsname\relax\def\natexlab#1{#1}\fi
\expandafter\ifx\csname bibnamefont\endcsname\relax
  \def\bibnamefont#1{#1}\fi
\expandafter\ifx\csname bibfnamefont\endcsname\relax
  \def\bibfnamefont#1{#1}\fi
\expandafter\ifx\csname citenamefont\endcsname\relax
  \def\citenamefont#1{#1}\fi
\expandafter\ifx\csname url\endcsname\relax
  \def\url#1{\texttt{#1}}\fi
\expandafter\ifx\csname urlprefix\endcsname\relax\def\urlprefix{URL }\fi
\providecommand{\bibinfo}[2]{#2}
\providecommand{\eprint}[2][]{\url{#2}}

\bibitem[{\citenamefont{Katsnelson}(2007)}]{katsnelson_MT_07}
\bibinfo{author}{\bibfnamefont{M.~I.} \bibnamefont{Katsnelson}},
  \bibinfo{journal}{Materials Today} \textbf{\bibinfo{volume}{10}},
  \bibinfo{pages}{20} (\bibinfo{year}{2007}).

\bibitem[{\citenamefont{Ijima}(1991)}]{ijima_nature_91}
\bibinfo{author}{\bibfnamefont{S.}~\bibnamefont{Ijima}},
  \bibinfo{journal}{Nature} \textbf{\bibinfo{volume}{354}}, \bibinfo{pages}{56}
  (\bibinfo{year}{1991}).

\bibitem[{\citenamefont{Baughman et~al.}(2002)\citenamefont{Baughman, Zakhidov,
  and {de Heer}}}]{baughman_science_02}
\bibinfo{author}{\bibfnamefont{R.~H.} \bibnamefont{Baughman}},
  \bibinfo{author}{\bibfnamefont{A.~A.} \bibnamefont{Zakhidov}},
  \bibnamefont{and} \bibinfo{author}{\bibfnamefont{W.~A.} \bibnamefont{{de
  Heer}}}, \bibinfo{journal}{Science} \textbf{\bibinfo{volume}{297}},
  \bibinfo{pages}{787} (\bibinfo{year}{2002}).

\bibitem[{\citenamefont{Kroto et~al.}(1985)\citenamefont{Kroto, Heath, O'Brien,
  Curl, and Smalley}}]{kroto_nature_85}
\bibinfo{author}{\bibfnamefont{H.~W.} \bibnamefont{Kroto}},
  \bibinfo{author}{\bibfnamefont{J.~R.} \bibnamefont{Heath}},
  \bibinfo{author}{\bibfnamefont{S.~C.} \bibnamefont{O'Brien}},
  \bibinfo{author}{\bibfnamefont{R.~F.} \bibnamefont{Curl}}, \bibnamefont{and}
  \bibinfo{author}{\bibfnamefont{R.~E.} \bibnamefont{Smalley}},
  \bibinfo{journal}{Nature} \textbf{\bibinfo{volume}{318}},
  \bibinfo{pages}{162} (\bibinfo{year}{1985}).

\bibitem[{\citenamefont{Meyer et~al.}(2007)\citenamefont{Meyer, Geim,
  Katsnelson, Novoselov, Booth, and Roth}}]{meyer_nature_07}
\bibinfo{author}{\bibfnamefont{J.~C.} \bibnamefont{Meyer}},
  \bibinfo{author}{\bibfnamefont{A.~K.} \bibnamefont{Geim}},
  \bibinfo{author}{\bibfnamefont{M.~I.} \bibnamefont{Katsnelson}},
  \bibinfo{author}{\bibfnamefont{K.~S.} \bibnamefont{Novoselov}},
  \bibinfo{author}{\bibfnamefont{T.~J.} \bibnamefont{Booth}}, \bibnamefont{and}
  \bibinfo{author}{\bibfnamefont{S.}~\bibnamefont{Roth}},
  \bibinfo{journal}{Nature} \textbf{\bibinfo{volume}{446}}, \bibinfo{pages}{60}
  (\bibinfo{year}{2007}).

\bibitem[{\citenamefont{Kinaret et~al.}(2003)\citenamefont{Kinaret, Nord, and
  Viefers}}]{kinaret_APL_03}
\bibinfo{author}{\bibfnamefont{J.~M.} \bibnamefont{Kinaret}},
  \bibinfo{author}{\bibfnamefont{T.}~\bibnamefont{Nord}}, \bibnamefont{and}
  \bibinfo{author}{\bibfnamefont{S.}~\bibnamefont{Viefers}},
  \bibinfo{journal}{Appl. Phys. Lett.} \textbf{\bibinfo{volume}{82}},
  \bibinfo{pages}{1287} (\bibinfo{year}{2003}).

\bibitem[{\citenamefont{Stankovich et~al.}(2006)\citenamefont{Stankovich,
  Dikin, Dommet, Kohlhaas, Zimney, Stach, Piner, Nguyen, and
  Ruoff}}]{stankovich_nature_06}
\bibinfo{author}{\bibfnamefont{S.}~\bibnamefont{Stankovich}},
  \bibinfo{author}{\bibfnamefont{D.~A.} \bibnamefont{Dikin}},
  \bibinfo{author}{\bibfnamefont{G.~H.~B.} \bibnamefont{Dommet}},
  \bibinfo{author}{\bibfnamefont{K.~M.} \bibnamefont{Kohlhaas}},
  \bibinfo{author}{\bibfnamefont{E.~J.} \bibnamefont{Zimney}},
  \bibinfo{author}{\bibfnamefont{E.~A.} \bibnamefont{Stach}},
  \bibinfo{author}{\bibfnamefont{R.~D.} \bibnamefont{Piner}},
  \bibinfo{author}{\bibfnamefont{S.~T.} \bibnamefont{Nguyen}},
  \bibnamefont{and} \bibinfo{author}{\bibfnamefont{R.~S.} \bibnamefont{Ruoff}},
  \bibinfo{journal}{Nature} \textbf{\bibinfo{volume}{442}},
  \bibinfo{pages}{282} (\bibinfo{year}{2006}).

\bibitem[{\citenamefont{Koskinen et~al.}(2007)\citenamefont{Koskinen,
  H{\"a}kkinen, Huber, {von Issendorff}, and Moseler}}]{koskinen_PRL_07}
\bibinfo{author}{\bibfnamefont{P.}~\bibnamefont{Koskinen}},
  \bibinfo{author}{\bibfnamefont{H.}~\bibnamefont{H{\"a}kkinen}},
  \bibinfo{author}{\bibfnamefont{B.}~\bibnamefont{Huber}},
  \bibinfo{author}{\bibfnamefont{B.}~\bibnamefont{{von Issendorff}}},
  \bibnamefont{and} \bibinfo{author}{\bibfnamefont{M.}~\bibnamefont{Moseler}},
  \bibinfo{journal}{Phys. Rev. Lett.} \textbf{\bibinfo{volume}{98}},
  \bibinfo{pages}{015701} (\bibinfo{year}{2007}).

\bibitem[{\citenamefont{Charlier et~al.}(1997)\citenamefont{Charlier, {De
  Vita}, Blase, and Car}}]{charlier_science_97}
\bibinfo{author}{\bibfnamefont{J.-C.} \bibnamefont{Charlier}},
  \bibinfo{author}{\bibfnamefont{A.}~\bibnamefont{{De Vita}}},
  \bibinfo{author}{\bibfnamefont{X.}~\bibnamefont{Blase}}, \bibnamefont{and}
  \bibinfo{author}{\bibfnamefont{R.}~\bibnamefont{Car}},
  \bibinfo{journal}{Science} \textbf{\bibinfo{volume}{275}},
  \bibinfo{pages}{647} (\bibinfo{year}{1997}).

\bibitem[{\citenamefont{Lee et~al.}(1997)\citenamefont{Lee, Kim, and
  Tomanek}}]{lee_PRL_97}
\bibinfo{author}{\bibfnamefont{Y.~H.} \bibnamefont{Lee}},
  \bibinfo{author}{\bibfnamefont{S.~G.} \bibnamefont{Kim}}, \bibnamefont{and}
  \bibinfo{author}{\bibfnamefont{D.}~\bibnamefont{Tomanek}},
  \bibinfo{journal}{Phys. Rev. Lett.} \textbf{\bibinfo{volume}{78}},
  \bibinfo{pages}{2393} (\bibinfo{year}{1997}).

\bibitem[{\citenamefont{Nakada et~al.}(1996)\citenamefont{Nakada, Fujita,
  Dresselhaus, and Dresselhaus}}]{nakada_PRB_96}
\bibinfo{author}{\bibfnamefont{K.}~\bibnamefont{Nakada}},
  \bibinfo{author}{\bibfnamefont{M.}~\bibnamefont{Fujita}},
  \bibinfo{author}{\bibfnamefont{G.}~\bibnamefont{Dresselhaus}},
  \bibnamefont{and} \bibinfo{author}{\bibfnamefont{M.~S.}
  \bibnamefont{Dresselhaus}}, \bibinfo{journal}{Phys. Rev. B}
  \textbf{\bibinfo{volume}{54}}, \bibinfo{pages}{17954} (\bibinfo{year}{1996}).

\bibitem[{\citenamefont{Kawai et~al.}(2000)\citenamefont{Kawai, Miyamoto,
  Sugino, and Koga}}]{kawai_PRB_00}
\bibinfo{author}{\bibfnamefont{T.}~\bibnamefont{Kawai}},
  \bibinfo{author}{\bibfnamefont{Y.}~\bibnamefont{Miyamoto}},
  \bibinfo{author}{\bibfnamefont{O.}~\bibnamefont{Sugino}}, \bibnamefont{and}
  \bibinfo{author}{\bibfnamefont{Y.}~\bibnamefont{Koga}},
  \bibinfo{journal}{Phys. Rev. B} \textbf{\bibinfo{volume}{62}},
  \bibinfo{pages}{R16349} (\bibinfo{year}{2000}).

\bibitem[{\citenamefont{Hernandez et~al.}(2000)\citenamefont{Hernandez,
  Ordejon, Boustani, Rubio, and Alonso}}]{hernandez_JCP_00}
\bibinfo{author}{\bibfnamefont{E.}~\bibnamefont{Hernandez}},
  \bibinfo{author}{\bibfnamefont{P.}~\bibnamefont{Ordejon}},
  \bibinfo{author}{\bibfnamefont{I.}~\bibnamefont{Boustani}},
  \bibinfo{author}{\bibfnamefont{A.}~\bibnamefont{Rubio}}, \bibnamefont{and}
  \bibinfo{author}{\bibfnamefont{J.~A.} \bibnamefont{Alonso}},
  \bibinfo{journal}{J. Chem. Phys.} \textbf{\bibinfo{volume}{113}},
  \bibinfo{pages}{3814} (\bibinfo{year}{2000}).

\bibitem[{\citenamefont{Jiang et~al.}(2007)\citenamefont{Jiang, Sumpter, and
  Dai}}]{jiang_JCP_07}
\bibinfo{author}{\bibfnamefont{D.}~\bibnamefont{Jiang}},
  \bibinfo{author}{\bibfnamefont{B.~G.} \bibnamefont{Sumpter}},
  \bibnamefont{and} \bibinfo{author}{\bibfnamefont{S.}~\bibnamefont{Dai}},
  \bibinfo{journal}{J. Chem. Phys.} \textbf{\bibinfo{volume}{126}},
  \bibinfo{pages}{134701} (\bibinfo{year}{2007}).

\bibitem[{\citenamefont{Gunlycke et~al.}(2007)\citenamefont{Gunlycke, Areshkin,
  and White}}]{gunlycke_APL_07}
\bibinfo{author}{\bibfnamefont{D.}~\bibnamefont{Gunlycke}},
  \bibinfo{author}{\bibfnamefont{D.~A.} \bibnamefont{Areshkin}},
  \bibnamefont{and} \bibinfo{author}{\bibfnamefont{C.~T.} \bibnamefont{White}},
  \bibinfo{journal}{Appl. Phys. Lett.} \textbf{\bibinfo{volume}{90}},
  \bibinfo{pages}{142104} (\bibinfo{year}{2007}).

\bibitem[{\citenamefont{Kobayashi et~al.}(2006)\citenamefont{Kobayashi, Fukui,
  Enoki, and Kusakabe}}]{kobayashi_PRB_06}
\bibinfo{author}{\bibfnamefont{Y.}~\bibnamefont{Kobayashi}},
  \bibinfo{author}{\bibfnamefont{K.-I.} \bibnamefont{Fukui}},
  \bibinfo{author}{\bibfnamefont{T.}~\bibnamefont{Enoki}}, \bibnamefont{and}
  \bibinfo{author}{\bibfnamefont{K.}~\bibnamefont{Kusakabe}},
  \bibinfo{journal}{Phys. Rev. B} \textbf{\bibinfo{volume}{73}},
  \bibinfo{pages}{125415} (\bibinfo{year}{2006}).

\bibitem[{\citenamefont{Jian et~al.}(2006)\citenamefont{Jian, Yan, Kulaots,
  Crawford, and Hurt}}]{jian_carbon_06}
\bibinfo{author}{\bibfnamefont{K.}~\bibnamefont{Jian}},
  \bibinfo{author}{\bibfnamefont{A.}~\bibnamefont{Yan}},
  \bibinfo{author}{\bibfnamefont{I.}~\bibnamefont{Kulaots}},
  \bibinfo{author}{\bibfnamefont{G.~P.} \bibnamefont{Crawford}},
  \bibnamefont{and} \bibinfo{author}{\bibfnamefont{R.}~\bibnamefont{Hurt}},
  \bibinfo{journal}{Carbon} \textbf{\bibinfo{volume}{44}},
  \bibinfo{pages}{2105} (\bibinfo{year}{2006}).

\bibitem[{\citenamefont{Porezag et~al.}(1995)\citenamefont{Porezag, Frauenheim,
  K\"ohler, Seifert, and Kaschner}}]{porezag_PRB_95}
\bibinfo{author}{\bibfnamefont{D.}~\bibnamefont{Porezag}},
  \bibinfo{author}{\bibfnamefont{T.}~\bibnamefont{Frauenheim}},
  \bibinfo{author}{\bibfnamefont{T.}~\bibnamefont{K\"ohler}},
  \bibinfo{author}{\bibfnamefont{G.}~\bibnamefont{Seifert}}, \bibnamefont{and}
  \bibinfo{author}{\bibfnamefont{R.}~\bibnamefont{Kaschner}},
  \bibinfo{journal}{Phys. Rev. B} \textbf{\bibinfo{volume}{51}},
  \bibinfo{pages}{12947} (\bibinfo{year}{1995}).

\bibitem[{\citenamefont{Stone and Wales}(1986)}]{stone_CPL_86}
\bibinfo{author}{\bibfnamefont{A.~J.} \bibnamefont{Stone}} \bibnamefont{and}
  \bibinfo{author}{\bibfnamefont{D.~J.} \bibnamefont{Wales}},
  \bibinfo{journal}{Chem. Phys. Lett.} \textbf{\bibinfo{volume}{128}},
  \bibinfo{pages}{501} (\bibinfo{year}{1986}).

\bibitem[{\citenamefont{Sha and Jackson}(2004)}]{sha_JACS_04}
\bibinfo{author}{\bibfnamefont{X.}~\bibnamefont{Sha}} \bibnamefont{and}
  \bibinfo{author}{\bibfnamefont{B.}~\bibnamefont{Jackson}},
  \bibinfo{journal}{J. Am. Chem. Soc.} \textbf{\bibinfo{volume}{126}},
  \bibinfo{pages}{13094} (\bibinfo{year}{2004}).

\bibitem[{\citenamefont{Hua et~al.}(2000)\citenamefont{Hua, Cagin, Che, and
  {Goddard III}}}]{hua_nanotech_00}
\bibinfo{author}{\bibfnamefont{X.}~\bibnamefont{Hua}},
  \bibinfo{author}{\bibfnamefont{T.}~\bibnamefont{Cagin}},
  \bibinfo{author}{\bibfnamefont{J.}~\bibnamefont{Che}}, \bibnamefont{and}
  \bibinfo{author}{\bibfnamefont{W.~A.} \bibnamefont{{Goddard III}}},
  \bibinfo{journal}{Nanotechnology} \textbf{\bibinfo{volume}{11}},
  \bibinfo{pages}{85} (\bibinfo{year}{2000}).

\bibitem[{\citenamefont{Kobayashi et~al.}(2005)\citenamefont{Kobayashi, Fukui,
  Enoki, Kusakabe, and Kaburagi}}]{kobayashi_PRB_05}
\bibinfo{author}{\bibfnamefont{Y.}~\bibnamefont{Kobayashi}},
  \bibinfo{author}{\bibfnamefont{K.-I.} \bibnamefont{Fukui}},
  \bibinfo{author}{\bibfnamefont{T.}~\bibnamefont{Enoki}},
  \bibinfo{author}{\bibfnamefont{K.}~\bibnamefont{Kusakabe}}, \bibnamefont{and}
  \bibinfo{author}{\bibfnamefont{Y.}~\bibnamefont{Kaburagi}},
  \bibinfo{journal}{Phys. Rev. B} \textbf{\bibinfo{volume}{71}},
  \bibinfo{pages}{193406} (\bibinfo{year}{2005}).

\bibitem[{\citenamefont{Fukunaga et~al.}(2001)\citenamefont{Fukunaga, Itoh,
  Orimo, Aoki, and Fujii}}]{fukunaga_JAC_01}
\bibinfo{author}{\bibfnamefont{T.}~\bibnamefont{Fukunaga}},
  \bibinfo{author}{\bibfnamefont{K.}~\bibnamefont{Itoh}},
  \bibinfo{author}{\bibfnamefont{S.}~\bibnamefont{Orimo}},
  \bibinfo{author}{\bibfnamefont{M.}~\bibnamefont{Aoki}}, \bibnamefont{and}
  \bibinfo{author}{\bibfnamefont{H.}~\bibnamefont{Fujii}}, \bibinfo{journal}{J.
  Alloys Compd.} \textbf{\bibinfo{volume}{327}}, \bibinfo{pages}{224}
  (\bibinfo{year}{2001}).

\bibitem[{\citenamefont{Li et~al.}(2008)\citenamefont{Li, Wang, Zhang, Lee, and
  Dai}}]{li_science_08}
\bibinfo{author}{\bibfnamefont{X.}~\bibnamefont{Li}},
  \bibinfo{author}{\bibfnamefont{X.}~\bibnamefont{Wang}},
  \bibinfo{author}{\bibfnamefont{L.}~\bibnamefont{Zhang}},
  \bibinfo{author}{\bibfnamefont{S.}~\bibnamefont{Lee}}, \bibnamefont{and}
  \bibinfo{author}{\bibfnamefont{H.}~\bibnamefont{Dai}},
  \bibinfo{journal}{Science} \textbf{\bibinfo{volume}{319}},
  \bibinfo{pages}{1229} (\bibinfo{year}{2008}).

\bibitem[{\citenamefont{Gan et~al.}(2008)\citenamefont{Gan, Kotakoski,
  Krasheninnikov, Nordlund, and Banhart}}]{gan_NJP_08}
\bibinfo{author}{\bibfnamefont{Y.}~\bibnamefont{Gan}},
  \bibinfo{author}{\bibfnamefont{J.}~\bibnamefont{Kotakoski}},
  \bibinfo{author}{\bibfnamefont{A.~V.} \bibnamefont{Krasheninnikov}},
  \bibinfo{author}{\bibfnamefont{K.}~\bibnamefont{Nordlund}}, \bibnamefont{and}
  \bibinfo{author}{\bibfnamefont{F.}~\bibnamefont{Banhart}},
  \bibinfo{journal}{New J. Phys.} \textbf{\bibinfo{volume}{10}},
  \bibinfo{pages}{023022} (\bibinfo{year}{2008}).

\bibitem[{\citenamefont{Perdew et~al.}(1996)\citenamefont{Perdew, Burke, and
  Ernzerhof}}]{perdew_PRL_96}
\bibinfo{author}{\bibfnamefont{J.~P.} \bibnamefont{Perdew}},
  \bibinfo{author}{\bibfnamefont{K.}~\bibnamefont{Burke}}, \bibnamefont{and}
  \bibinfo{author}{\bibfnamefont{M.}~\bibnamefont{Ernzerhof}},
  \bibinfo{journal}{Phys. Rev. Lett.} \textbf{\bibinfo{volume}{77}},
  \bibinfo{pages}{3865} (\bibinfo{year}{1996}).

\bibitem[{\citenamefont{Bl\"ochl}(1994)}]{blochl_PRB_94}
\bibinfo{author}{\bibfnamefont{P.~E.} \bibnamefont{Bl\"ochl}},
  \bibinfo{journal}{Phys. Rev. B} \textbf{\bibinfo{volume}{50}},
  \bibinfo{pages}{17953} (\bibinfo{year}{1994}).

\bibitem[{gpa()}]{gpaw_wiki}
\emph{\bibinfo{title}{{GPAW} wiki}},
  \urlprefix\url{https://wiki.fysik.dtu.dk/gpaw}.

\bibitem[{\citenamefont{Henkelman and Jonsson}(2000)}]{henkelman_JCP_00}
\bibinfo{author}{\bibfnamefont{G.}~\bibnamefont{Henkelman}} \bibnamefont{and}
  \bibinfo{author}{\bibfnamefont{H.}~\bibnamefont{Jonsson}},
  \bibinfo{journal}{J. Chem. Phys.} \textbf{\bibinfo{volume}{113}},
  \bibinfo{pages}{9978} (\bibinfo{year}{2000}).

\bibitem[{\citenamefont{Bitzek et~al.}(2006)\citenamefont{Bitzek, Koskinen,
  G\"ahler, Moseler, and Gumbsh}}]{bitzek_PRL_06}
\bibinfo{author}{\bibfnamefont{E.}~\bibnamefont{Bitzek}},
  \bibinfo{author}{\bibfnamefont{P.}~\bibnamefont{Koskinen}},
  \bibinfo{author}{\bibfnamefont{F.}~\bibnamefont{G\"ahler}},
  \bibinfo{author}{\bibfnamefont{M.}~\bibnamefont{Moseler}}, \bibnamefont{and}
  \bibinfo{author}{\bibfnamefont{P.}~\bibnamefont{Gumbsh}},
  \bibinfo{journal}{Phys. Rev. Lett.} \textbf{\bibinfo{volume}{97}},
  \bibinfo{pages}{170201} (\bibinfo{year}{2006}).

\end{thebibliography}
\end{document}